\documentclass[10pt]{article}
\usepackage[OE]{express}
\usepackage{graphicx}
\usepackage{subfigure}
\usepackage{braket}
 \usepackage{color}
\usepackage{ragged2e}
  \definecolor{Revise1}{rgb}{0.0,0.0,0.0}
 \definecolor{Revise2}{rgb}{0.0,0.0,0.0}
\begin{document}
\title{Coherent  frequency bridge between visible and telecommunications band for vortex light}
\author{
\raggedright
Shi-Long Liu,\authormark{1,2} Shi-Kai Liu,\authormark{1,2} Yin-Hai Li,\authormark{1,2} Shuai-Shi,\authormark{1,2} Zhi-Yuan Zhou,\authormark{1,2,*} and Bao-Sen Shi\authormark{1,2}}
\address{\authormark{1}Key Laboratory of Quantum Information, University of Science and Technology of China, Hefei, Anhui 230026, China\\
\authormark{2}Synergetic Innovation Center of Quantum Information \& Quantum Physics, University of Science and Technology of China, Hefei, Anhui 230026, China\\
\email{\authormark{*}zyzhouphy@ustc.edu.cn} }%

\begin{abstract}
In quantum communications, vortex photons can encode higher-dimensional quantum states and build high-dimensional communication networks (HDCNs). The interfaces that connect different wavelengths are significant in HDCNs. We construct a coherent orbital angular momentum (OAM) frequency bridge via difference frequency conversion in a nonlinear bulk crystal for HDCNs. {\color{Revise2}{Using a single resonant cavity, maximum quantum conversion efficiencies from visible to infrared are 36\%, 15\%, and 7.8\% for topological charges of 0,1, and 2, respectively.}} The average fidelity obtained using quantum state tomography for the down-converted infrared OAM-state of topological charge 1 is 96.51\%. We also prove that the OAM is {\color{Revise2}{conserved}} in this process by measuring visible and infrared interference patterns.
This coherent OAM frequency-down conversion bridge represents a basis for an interface between two high-dimensional quantum systems operating with different spectra.
\end{abstract}

\section{Introduction}
 Orbital angular momentum (OAM),  a remarkable photonic freedom that inherently has multiple dimensions, can enhance the information capacity in communications. The pioneering work in 1992\cite{Allen1992} of Allen \emph{et al}. demonstrated that light beams with azimuthal phase dependence of exp($iL\phi$) can carry OAM of $L\hbar$. {\color{Revise2}{These vortex beams have interesting characteristics that have stimulated major research interest in several fields, including optical manipulation and trapping\cite{Padgett2011,Paterson2001,Yao2011}, high-capacity communications\cite{Wang2012,Willner2015}, spiral imaging, and high-precision optical measurements\cite{Chen2014,Torner2005,Zhou2014}.}} In quantum information fields, photons can be encoded in the OAM space because of the basis of OAM with its infinite dimensions in Hilbert space, which can then generate high-dimensional entanglement states\cite{Dada2011, Mair2001} and can be used to build OAM quantum memory for high-dimensional communication networks (HDCN)\cite{Vallone2014,Ding2013,Nicolas2014,Ding2015}. To date, most atom-based quantum repeaters operate in the visible spectrum. However, fiber-based quantum communication networks generally work in the telecommunications band, which lies within the low-loss communication windows \cite{Duan2001,simon2003robust,Bussieres2014}. Therefore, a coherent frequency bridge is required to connect the two spectra. In 1990, Kumar \emph{et al.} first proposed the concept of a quantum frequency converter using second-order nonlinearity to change the frequency of the quantum state while maintaining other quantum properties\cite{kumar1990quantum}. Subsequently, many quantum frequency converters have been constructed to connect different systems\cite{Huang1992,Albota2004,Rakher2010,Zaske2012,vollmer2014quantum}.
 Two reversible processes are usually required to realize the interface between visible and telecommunications-band photons: quantum frequency up-conversion (QFUC) and quantum frequency down-conversion (QFDC).  QFUC, which is based on sum frequency generation (SFG), can convert infrared photons into visible photons. Many previous researchers have realized QFUC experimentally on the single photon level in different material systems, including periodically poled lithium niobate (PPLN) and potassium titanyl phosphate (PPKTP) nonlinear crystals and PPLN waveguides\cite{Albota2004,Rakher2010,Ramelow2012,Rutz2017,zhou2017super}.
 In contrast to up-conversion, the reverse process of QFDC, which is based on difference frequency generation, has been rapidly developed in recent years \cite{Goldberg1995,curtz2010coherent,takesue2010single,Zaske2012,ikuta2011wide}.
There are two possible configurations for frequency conversion of OAM modes: one is the single pass configuration; the other is the cavity-enhanced single resonance scheme. In the first configuration, the two input fields make single passes through the crystal. Shao \emph{et al}. presented a theoretical model and a simulation of OAM frequency conversion using a quasi-phase-matching (QPM) crystal\cite{shao2013nonlinear}. Later, Li \emph{et al}. provided an analytical expression for frequency conversion in the SFG process\cite{Li2015}.  Steinlechner \emph{et al.} proposed an SFG-based conversion scheme to convert structured light from the near infrared (803 nm) to the visible range (527 nm) \cite{steinlechner2016frequency}, although the power conversion efficiency achieved was low (0.09\% for L=1). The second cavity-enhanced configuration can increase the conversion efficiency obviously when compared with the first scheme, where the basic Gaussian pump beam is resonant with the cavity and the input photons make single passes through the crystal. Using the second setup, Zhou \emph{et al}. realized frequency up-conversion of OAM modes from the infrared to the visible band based on SFG\cite{Zhou2016}, in which the maximum conversion efficiency of the OAM photon for topological charge 1 was 8.3\%. However, down-conversion of OAM modes has not been reported to date. Additionally, because superconducting detectors have near-unity quantum efficiency at 1550 nm, the down-conversion process has the potential to become much more useful.

In this work, we present the first demonstration of down-conversion for the OAM mode from a visible laser operating at 525 nm to an infrared laser beam at 1550 nm using a strong pump beam at 794 nm. Based on the nonlinear coupling equation, we propose an analytical expression to describe the conversion efficiency of the OAM down-conversion frequency bridge. The single-pass efficiencies realized for conversion from visible to infrared are 1.6\%, 0.52\%, and 0.4\% for L=0, 1, and 2, respectively, and the corresponding maximum quantum conversion efficiencies are 36\%, 15\%, and 7.8\%, respectively. {\color{Revise2}{In order to test the coherence of the converted OAM photons}}, we analyze the density matrix of the infrared OAM photons using quantum tomography\cite{James2001}. The average output OAM photon fidelity for topological charge 1 is 96.18\%. The high fidelity and high conversion efficiency indicate that the OAM down-conversion frequency bridge is both reliable and useful, which will pave the way for use of this bridge at the interface between two high-dimensional quantum systems with different spectra.

%
%
\section{Principle}
Difference frequency generation (DFG), which is based on a second-order nonlinearity, involves mixing of three waves such that they interact in a quasi-phase-matching periodically {\color{Revise2}{poled}} nonlinear crystal. During this process, the energy $(\omega_v-\omega_p=\omega_i)$, momentum $(k_v-k_p-k_i-2\pi/\Lambda=0)$, and OAM $(L_v-L_p-L_i=0)$ must be conserved, regardless of the forms of the interacting fields. Here, $k_v, k_p,$ and $k_i$ represent the properties of the input fundamental visible laser, the strong pump laser and the output infrared laser, respectively. In the DFG-based down-conversion experiment, the basic strong pump laser $P_p$ has a Gaussian beam, and the output infrared laser $P_i$ is dependent on the profile of the input visible beam $P_v$. By ignoring the time components of the strong Gaussian field, the spatial field can be written as\cite{Broyer1985}:
\begin{equation}\label{1}
  E_p(r,z)=\sqrt{\frac{P_p}{\pi\epsilon_0n_pc}
}\frac{1}{w_{0p}(1+i\tau_p)}exp\left(-\frac{r^2}{w_{0p}^2(1+i\tau_p)}\right)
\end{equation}
Here, $w_{0p}$ is the beam waist; $\tau_{p} (=2z/b_p)$ is a variable related to the propagation distance; $b_p (=k_pw_{0p}^2)$ is the confocal parameter of the Gaussian beam; and $n_p$ is the refractive index of the pump laser.  The input visible OAM beam has a similar expression with the exception of the topological charge L\cite{Zhou2016}:
\begin{equation}\label{2}
  E_v(r,z,L)=\sqrt{\frac{P_v}{\pi\epsilon_0n_vcL!}
}\frac{\left(\sqrt2r\right)^2}{w_{0v}(1+i\tau_v)}exp\left(-\frac{r^2}{w_{0v}^2(1+i\tau_v)}+iL\phi\right)
\end{equation}
   Based on the nonlinear coupled equation, the evolution of the output fields can be written as\cite{boyd1968parametric}:
\begin{equation}\label{3}
  \frac{dE_i}{dz}=K_iE_p^{*}E_ve^{-i\Delta kz}
\end{equation}
where $K_i (=2id_{eff}\omega_i/n_ic)$ is the coupling coefficient of the infrared laser. $d_{eff}$ is the effective nonlinear efficiency of the crystal. The two beams are assumed to be focused at the crystal$'$s center at $z=0$. On the output surface of the crystal, we obtain the output power as follows:
\begin{equation}\label{4}
  P_i(L/2)=2\epsilon_0cn_i\int E_iE_i^{*}ds=\frac{16\pi^2d_{eff}^2l2^L}{\epsilon_0cn_vn_i\lambda_i^2\lambda_p}
  h(\alpha,\beta,\xi,\sigma)\cdot P_pP_v
\end{equation}
where $n_v$ and $n_i$ are the refractive indices of the visible ($\lambda_v$) and infrared laser ($\lambda_i$) in crystal, respectively; $l$ is the length of the crystal; and $h(\alpha,\beta,\xi,\sigma)$ is the focusing function that is determined based on the waist ratio of the two beams:
\begin{equation}\label{5}
  h=\frac{1}{\xi}\int_{-\xi}^{\xi}\int_{-\xi}^{\xi}\frac{(1-ix)^{-1}(1+iy)^{-1}e^{-i\sigma(x-y)}}{(\alpha\beta^2(\beta+ix)(\beta-iy)(\frac{1}{1-ix}+\frac{1}{1+iy})
  +\beta^3(2\beta-iy+ix))^{L+1}}dxdy
\end{equation}
  Here, $\alpha (=w_{0v}^2/w_{0p}^2)$ and $\beta (=b_v/b_p)$ are the ratio of the squares of the beam waist and confocal parameters, respectively; and $\xi (=l/b_p)$ and $\sigma (=\Delta kb_p/2)$ are the pump focusing and spatial phase-mismatching parameters, respectively. When the passive transmission loss and the crystal's absorption $\delta(=T_{in}^vexp(-\alpha _Ll)T_{out}^i)$\cite{roussev2004periodically} are considered, Eq. (4) can be simplified to read:
\begin{equation}\label{6}
  P_i=K_Lh\delta\cdot P_pP_v
\end{equation}
Here, $K_L$ represents a constant with the topological charge and the parameters of the crystal that were used in Eq. (4). Based on Eq. (6), the single pass conversion efficiency (SPCE), the quantum conversion efficiency (QCE) and the maximum power of the strong pump power ($P_{p\_Max}$) for unit one conversion  can be calculated. The QCE from the visible spectrum to the infrared is defined as\cite{Albota2004}:
\begin{equation}\label{7}
  \eta=N_i/N_v=sin(\pi/2\sqrt{P_p/P_{p\_Max}})
\end{equation}
where $P_{p\_Max} (=\lambda_v/\lambda_iK_Lh(\alpha,\beta,\xi,\sigma)\delta)$ is the maximum pump power. Because of the high divergence angle for the higher-order OAM mode\cite{padgett2015divergence}, we increase the pump power for a higher conversion efficiency. In this down-conversion experiment, a resonant cavity is used, as shown in Fig. 1(a). When the cavity length is locked to the master laser, the circulating power can then be written approximately as $P_{circ}=F/2\pi\cdot P_p$, where F is the {\color{Revise2}{finesse}} of the cavity. In Fig. 1(b), we plot the theoretical SPCE and $P_{p\_Max}$ for the different input topological charges L. Here, the x-axis represents L, ranging from 0 to 10, and the left and right y-axes are the SPCE and $P_{p\_Max}$ with logarithmic scales, respectively. In our calculations, the waist sizes of the two beams are $w_p=60\ \mu m$ and $w_v (=50\times \sqrt{L+1}\ \mu m)$. The crystal used is a bulk PPLN crystal, for which the passive loss $\delta=0.95^3$. From Fig. 1(b), we find that the relationships of t he SPCE and $P_{p\_Max}$ with L are both approximately quadratic. For L=0 (Gaussian beam), L=1, and L=2, $P_{p\_Max}$ has values of 20 W, 50 W, and 166 W, respectively. For higher-order OAM modes, the values of $P_{p\_Max}$ are numerous.
\begin{figure}[!htbp]
  \centering
  \includegraphics[width=12.0cm]{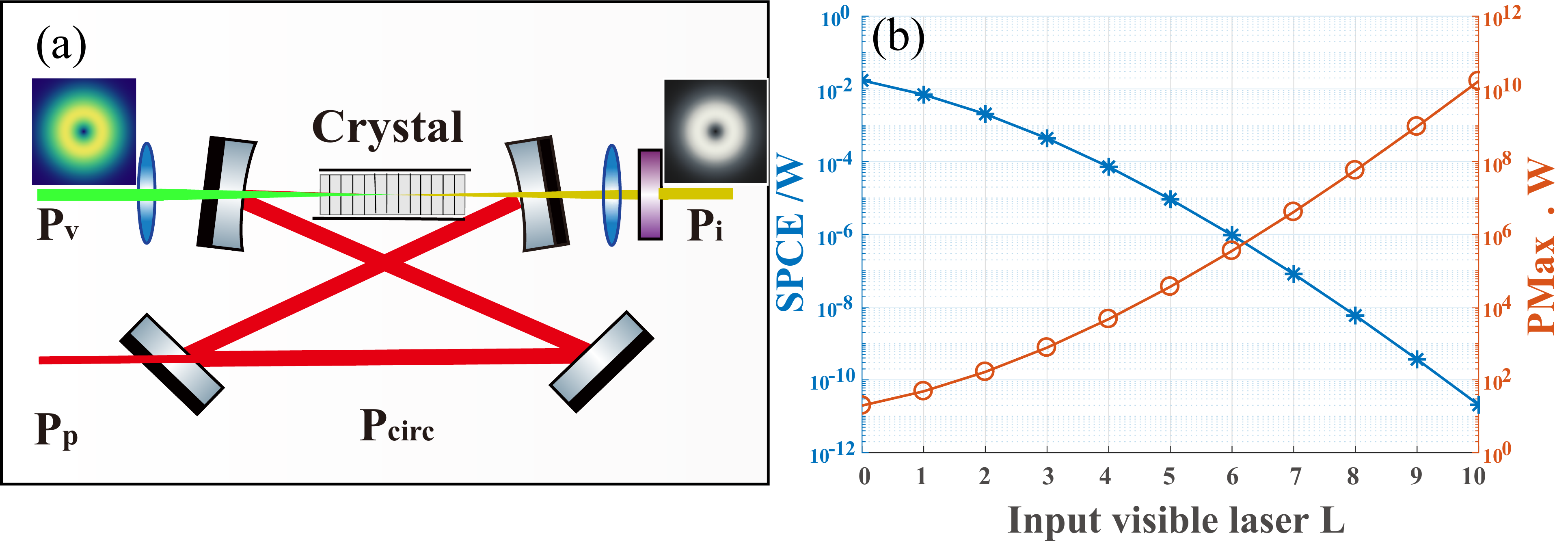}
  \caption{OAM coherent frequency bridge and theory model. (a) Simple setup for quantum frequency down-conversion (QFDC), where $P_p$, $P_v$, and $P_i$ are the pump, visible and infrared laser beam powers, respectively. $P_{circ}$ is the intra-cavity power. (b) Relationships of the SPCE (left y-axis) and $P_{p\_Max}$ (right y-axis) with the topological charge L of the input visible fields. The y-axis scales are logarithmic.}\label{1}
\end{figure}
\section{Experiment}
The design and testing of the coherent OAM down-frequency bridge are introduced in this section, including the {\color{Revise2}{detailed}} experimental setup and the conversion results.
\subsection{Experimental setup}
 A detailed schematic of the {\color{Revise2}{difference frequency generation (DFG)}} process is shown in Fig. 2. There are four modules: the photon source, state preparation, state conversion and state tomography modules.
 \begin{figure}[!htbp]
   \centering
   \includegraphics[width=12cm]{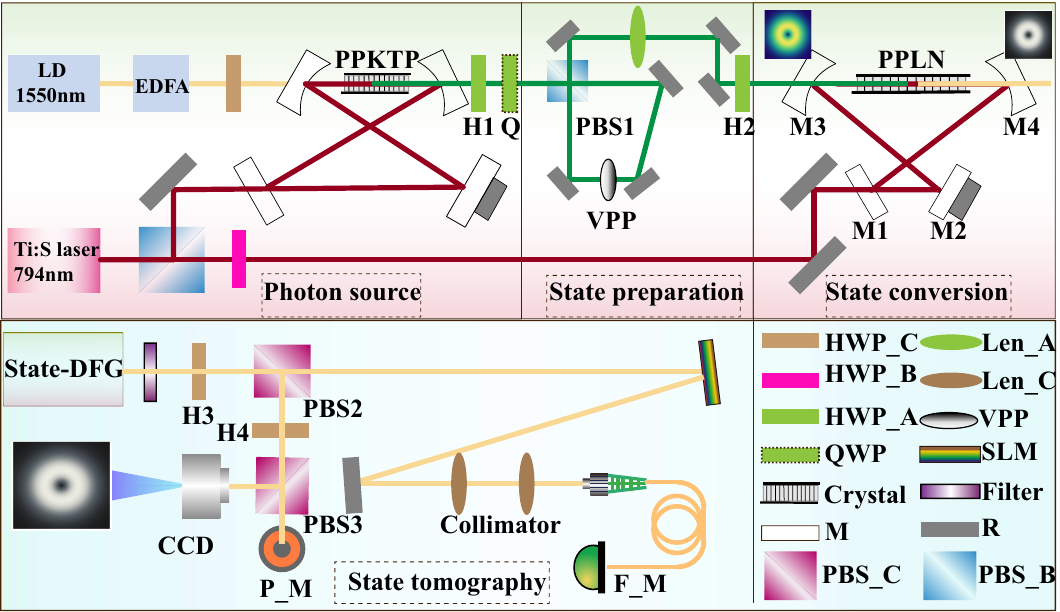}
   \caption{Setup for DFG-based OAM down-conversion. LD: diode laser (Toptica, predesigned ,1520 nm-1590 nm); EDFA: Er-doped fiber amplifier (1540 nm-1560 nm); Ti: sapphire laser (Coherent, MBR110); HWP: half-wave plate; QWP: quarter-wave plate; PBS\_B(C): polarizing beam splitter; VPP: vortex phase plate; Filter: long pass filter; SLM: infrared spatial light modulator; P\_M: power meter; F\_M: fiber power meter; PPKTP: (Raycol QPM crystals ; period of 9.375 $\mu m$); PPLN: {\color{Revise2}{(HCP, periodically poled lithium niobate (PPLN) chip SFVIS-MA; period of 7.3 $\mu m$)}}.}\label{2}
 \end{figure}

 The first setup shown in Fig. 2 is that of the photon source. The 794 nm laser beam comes from a Ti: sapphire laser, which is resonant with the cavity to produce strong pump power. The seeded 1550 nm infrared laser, which comes from a diode laser, is amplified via the EDFA and a single pass through the PPKTP crystal. When the quasi-phase-matching (QPM) condition is satisfied for the two fundamental laser beams in PPKTP, a high-quality 525 nm Gaussian beam is generated.

 The next part shows the state preparation structure. A modified Sagnac interferometer is used to produce an arbitrary OAM state, and includes a vortex phase plate, wave plates (H1, Q), and a polarizing beam splitter. After the interferometer, the two beams acquire the opposite and equal spatial vortex phases $exp({iL\phi})$, and the superposition state generated can be written as\cite{Zhou2016}:
 \begin{equation}\label{8}
   \ket{\phi}=1/\sqrt{2}\cdot(\ket{h,L}+e^{i\theta}\ket{v,-L})
 \end{equation}
  where $h,v$ represent the horizontal and vertical polarizations of the input state, respectively. In particular, when the optical axes of H2 are rotated by $22.5^{\circ}$ along the horizontal direction, the preparation state can be expressed as:
 \begin{equation}\label{9}
 \ket{\phi}_{pre}=1/2\cdot((\ket{L}+e^{i\theta}\ket{-L})\ket{h}+(\ket{L}+e^{i(\theta+\pi)}\ket{-L})\ket{v})
 \end{equation}

 The next section is the state down-conversion design, which is crucial to realization of down-conversion of the OAM mode from a {\color{Revise2}{visible laser (VL) beam to an infrared laser (IL) beam}}. This section consists of a single resonant cavity and a nonlinear crystal. The cavity parameters are optimized based on the theory of Boyd and Kleinman \cite{boyd1968parametric}. We select the focusing parameters $\xi=0.93,\mu=0.66$, and the corresponding beam waist of the 794 nm pump beam is 56 $\mu m$ at the center of the PPLN. {\color{Revise2}{The PPLN that was used in this experiment with a type-0 (e+e->e) quasi-phase-matching (QPM) condition for DFG has a cross-section of 50$\times$7.9$\times$0.5 $ mm^3$ (L$\times$ W$\times$ T), and was placed within a homemade oven. The selected period of the PPLN is 7.30 $\mu m$.}} The ring cavity has four mirrors, where M1 is the input mirror with 97\% reflection at 794 nm, and the plane mirror M2 and the two curved mirrors M3 and M4, with their 80 $mm$ radius of curvature , have high-reflection($>99.8\%$) coating for 794 nm. The crystal surface and the two curved mirrors have anti-reflection ($<1\%$) coating for the two laser beams. By measuring the leaked power after M2, we can then estimate the power that is circulating in the cavity.

   In order to test the coherence of the output infrared OAM state, we use two strategies. On the one hand, the shapes and the power of the infrared fields are observed using an infrared charge-coupled device (CCD) and a power meter. On the other hand, we calculate the density matrix and its fidelity using quantum state tomography\cite{James2001,Nicolas2015,zhou2016orbital}. The infrared light is filtered using a long pass filter, projected onto the spatial light modulator (SLM), and then collected into the fiber using collimators. The four projections selected on the basis of the SLM are $\ket{R}$, $\ket{L}$, $\ket{H}$,and $\ket{A}$, where $\ket{R}$ and $\ket{L}$ represent the eigenstates in self-representation for the OAM, and $\ket{H}=1/\sqrt{2}\cdot(\ket{R}+\ket{L})$ and $\ket{A}=1/\sqrt{2}\cdot(\ket{R}-i\ket{L})$
 are the supposition state.
 \subsection{DFG-based OAM quantum-frequency down-conversion}
 \begin{figure}[!htbp]
   \centering
   \includegraphics[width=12cm]{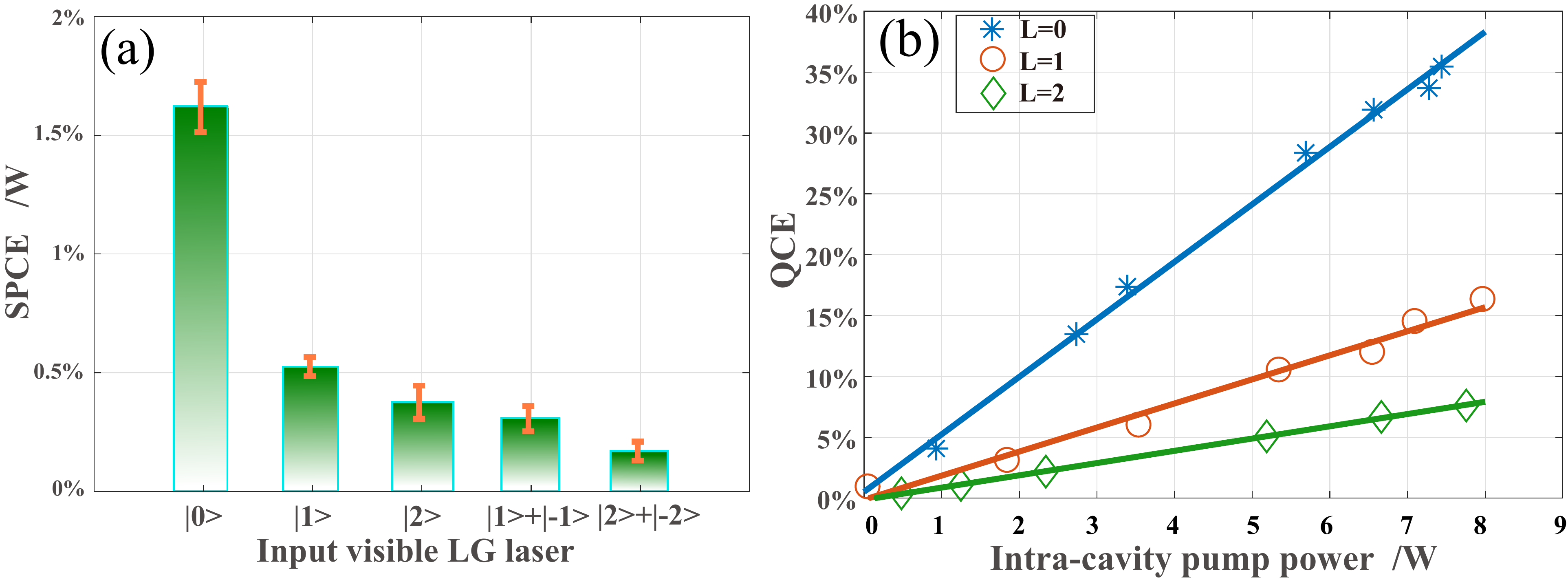}
   \caption{{\color{Revise2}{Single pass conversion efficiency (SPCE) and quantum conversion efficiency (QCE)}} of the OAM frequency bridge. (a) SPCE of the OAM state. The green bar and the red error bars represent the experimental and error values, respectively. The input fields are the Gaussian field ($\ket{0}$), a pure LG state  ($\ket{1},\ket{2}$), and the supposition state ($\ket{1}+\ket{-1}$, $\ket{2}+\ket{-2}$). (b) QCE for conversion from visible to infrared for L=0, 1, and 2 in the ring cavity. The x-axis represents the measured intra-cavity pump power.
   }\label{3}
 \end{figure}
 First, we measured the single pass conversion efficiency (SPCE) from the visible range to the infrared in frequency down-conversion of the OAM mode, where the two laser beams make a single pass through the PPLN. The pump laser waist is approximately 60 $\mu m$, and the input laser is different green OAM state. The optimal temperature is $29.50^{\circ } C$. The main results are shown in Fig. 3(a), where the x-axis represents the input fields, including the Gaussian ($\ket{0}$), the pure OAM state ($\ket{1}, \ket{2}$, as shown in Eq. (8)), and the supposition state ($\ket{1}+\ket{-1}$, $\ket{2}+\ket{-2}$, as shown in Eq. (9)). The y-axis is the measured SPCE. The green bars are the measured average values of the SPCE. In this process, the visible power is fixed at 40 mW, and the pump power is increased linearly from 50 mW to 500 mW. Because of system jitter, error bars must be inserted on top of the data, {\color{Revise2}{and these error bars represent the standard deviation in a group of measurements}}. {\color{Revise1}{Because of the type-0(e+e->) quasi-phase-matching condition, the conversion efficiency of the supposition state ($\ket{1}+\ket{-1}$;$\ket{2}+\ket{-2}$) is half of that of the corresponding pure state ($\ket{1}$;$\ket{2}$), based on Eq. (9)}}. The conversion efficiency is lower than the theoretically predicted figure shown in Fig. 1(b), which is based on the uncertainty of the beam waist and the imperfect mode overlap. From Fig. 3(a),  we find that the conversion efficiency for higher-order OAM modes is very weak in the single-pass configuration. To obtain high conversion efficiency, we place the crystal within a single resonant ring cavity.

 Second, we present the main results of the cavity-enhanced configuration in Fig. 3(b), which shows that the intra-cavity power is a function of the circulated pump power. Here, the x-axis represents the measured intra-cavity power, and the y-axis is the intra-cavity {\color{Revise2}{quantum conversion efficiency (QCE)}}. The data represented by blue asterisks, red circles, and green diamonds are the QCEs for L=0, L=1, and L=2, respectively, and the three corresponding lines are the fitting results based on the least squares method. To estimate the intra-cavity QCE, linear attenuation factors of approximately $92\%$ are taken into account. Here, the {\color{Revise2}{finesse}} of the cavity for the pump laser is 85, and thus the enhanced factor is approximately 14 when the cavity is locked onto the master laser. Under lower pump powers, the relationship between the QCE and circulated power show good linearity. For the Gaussian beam, unit conversion efficiency can be reached when the intra-cavity power increases by three times with respect to the current maximum power, which is the same as the theoretical prediction given in Fig. 1(b).
 \begin{figure}[!htbp]
  \centering
  \includegraphics[width=10cm]{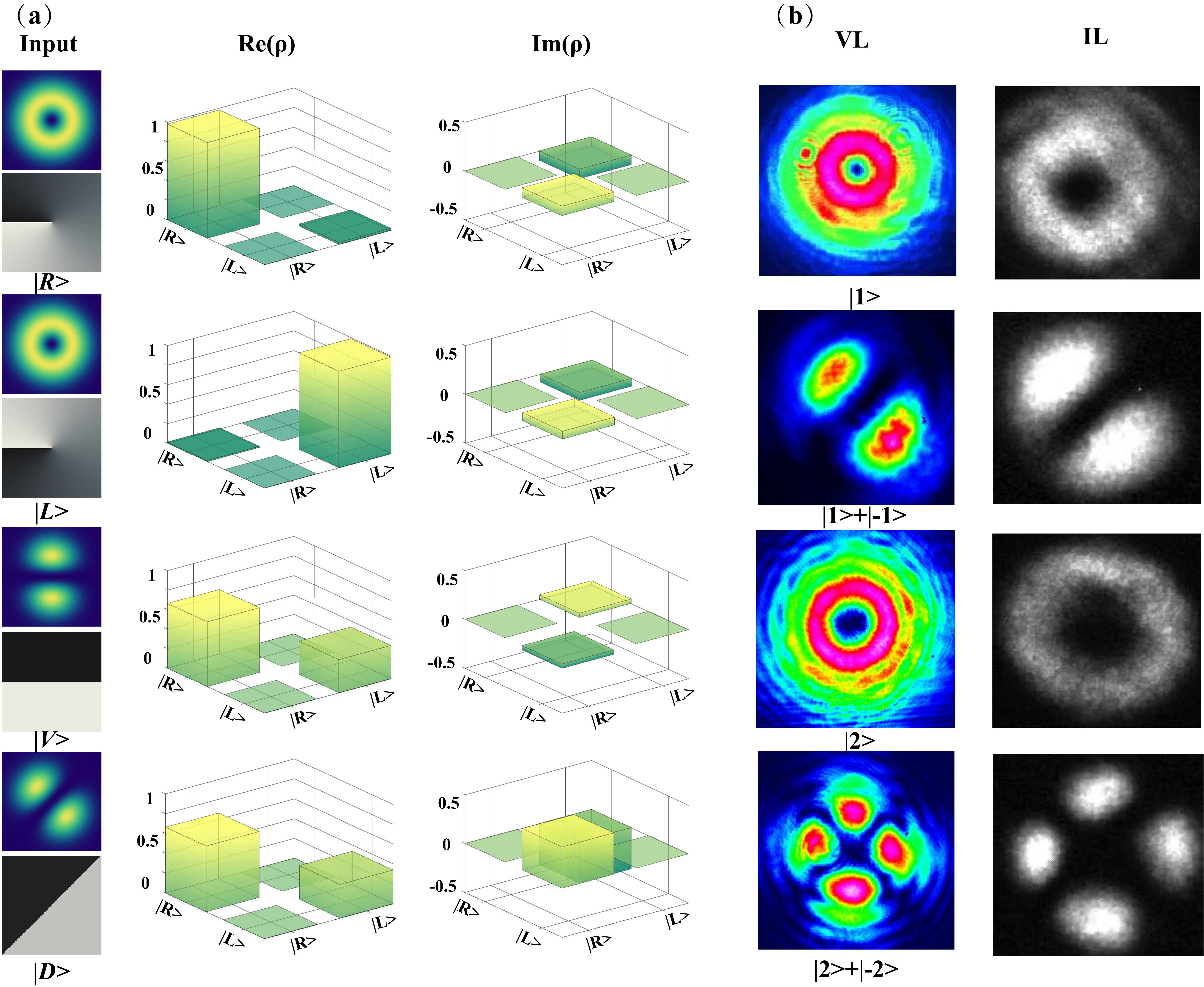}
  \caption{Density matrix and intensity profiles. (a) Input state and density matrix of the output state. The input row describes the intensity and the phase of the input fields; the lists of $Re(\rho)$ and $Im(\rho)$ are the real and imaginary parts of the density matrix for the output fields determined using quantum state tomography. (b) Intensity profiles of the input {\color{Revise2}{visible laser (VL) and the output infrared laser (IL)}} beams determined using visible and infrared CCDs.}
  \end{figure}

 Finally, we test the fidelity of the output infrared OAM-mode. The 3D graph shown in Fig. 4(a) represents the density matrix of the infrared output state for topological charge 1 that are determined by quantum state tomography, where the components on the left are the intensities and phase distributions of the input infrared OAM fields, and the components on the right are the corresponding density matrices. In this process, the input states are $\ket{R}$, $\ket{L}$, $\ket{V}$, and  $\ket{D}$. For each input state, the four corresponding projection bases that are loaded on the measured SLM are $(\ket{R}, \ket{L}, \ket{A},$ and $\ket{H})$. The average fidelities $\bra{\phi}\rho\ket{\phi}$ of $\ket{R}$, $\ket{L}$, $\ket{V}$, and $\ket{D}$ are 98.01\%, 98.82\%, 97.01\% and 92.23\%, respectively. The fidelity of $\ket{D}$ is lower than that of the other input fields, which represents the imperfect preparation of the input state and the deflected positions of the projection {\color{Revise2}{bases}} in the SLM. Nevertheless, the high fidelity and the high conversion efficiency for down-conversion of the OAM mode indicate the reliable performance of the method. Without generality, we show the intensity profiles of the two fields in Fig. 4(b), as measured using visible and infrared CCDs. The colorized graphs shown on the left of Fig. 4(b) are the input pure or superposition states $\ket{1}$, $\ket{1}+\ket{-1}$, $\ket{2}$, and $\ket{2}+\ket{-2}$, and the corresponding gray graphs on the right are the corresponding infrared OAM modes. It can been seen that the intensity profiles of the input and output fields are highly similar. Based on observation of the interference of the superposition state, we can also determine that the topological charge of the output state is equal to that of the input state, i.e., the OAM is conserved in this process.
\section{Conclusion}
In summary, we have demonstrated coherent OAM frequency down-conversion from the visible range to the infrared based on difference frequency generation in a single resonant cavity. We first propose a theoretical model to describe the OAM frequency down-conversion process. Then, we demonstrate OAM frequency down-conversion experimentally for different OAM modes. The maximum quantum conversion efficiencies that were obtained for OAM modes with topological charges of 0, 1, and 2 were 36\%, 15\%, and 7.8\%, respectively. Using quantum state tomography, the average fidelity achieved was determined to be 96.5\%. We also showed that the OAM is conserved and that the coherent property is preserved during the DFG process. The high fidelity and high QCE values obtained show that our OAM frequency-down converter is reliable and has potential applications in construction of high-dimensional quantum networks. When compared with recently reported results in OAM frequency up-conversion, our OAM frequency down-conversion process has shown higher conversion efficiencies\cite{Li2015,steinlechner2016frequency,zhou2016orbital}.

 The remaining problem that must be solved is further enhancement of the quantum conversion efficiency for the higher-order modes. Use of strong pump laser is one possible way, but the best method would involve realization of mode-independent conversion efficiency for the higher-order modes with a fixed pump power. This issue will be investigated in our future research. When this mode-independent conversion efficiency is realized, frequency conversion of higher-dimensional OAM states would then be feasible. We will study frequency down-conversion of the single photon OAM states, two-dimensional OAM entangled states and high-dimensional OAM-entanglement states in the near future. Because of the near optimal quantum detection efficiency of superconducting detectors operating at 1550 nm, this frequency conversion process will be very useful in the frequency conversion detection of visible, mid-infrared and far infrared light beams \cite{ikuta2011wide,mancinelli2017mid}.

\section{Acknowledgments}
We would like to thank Yu Zhang of Hangzhou Normal University for help with the experiments.
National Natural Science Foundation of China (11174271, 11604322, 61275115, 61435011, 61525504, 61605194); China Postdoctoral Science Foundation (2016M590570); Fundamental Research Funds for the Central Universities. We thank David MacDonald, MSc, from Liwen Bianji, Edanz Group China (www.liwenbianji.cn/ac), for editing the English text of a draft of this manuscript.
\end{document}